\documentclass{INTERSPEECH2023}

\usepackage{tipa}
\usepackage{graphicx}

\interspeechcameraready

\title{Speaker-independent Speech Inversion for Estimation of Nasalance}
\name{Yashish M. Siriwardena$^1$, Carol Espy-Wilson$^1$, Suzanne Boyce$^2$, Mark K.Tiede$^3$, Liran Oren$^2$}
\address{
  $^1$University of Maryland College Park, Maryland, USA\\
  $^2$University of Cincinnati, Ohio, USA \\
  $^3$Haskins Laboratories, Connecticut, USA
  }
\email{yashish@umd.edu, espy@umd.edu, boycese@ucmail.uc.edu, mark.tiede@yale.edu, orenl@ucmail.uc.edu}

\begin{document}

\maketitle

\begin{abstract}

The velopharyngeal (VP) valve regulates the opening between the nasal and oral cavities. This valve opens and closes through a coordinated motion of the velum and pharyngeal walls. Nasalance is an objective measure derived from the oral and nasal acoustic signals that correlate with nasality. In this work, we evaluate the degree to which the nasalance measure reflects fine-grained patterns of VP movement by comparison with simultaneously collected direct measures of VP opening using high-speed nasopharyngoscopy (HSN). We show that nasalance is significantly correlated with the HSN signal, and that both match expected patterns of nasality. We then train a temporal convolution-based speech inversion system in a speaker-independent fashion to estimate VP movement for nasality, using nasalance as the ground truth. In further experiments, we also show the importance of incorporating source features (from glottal activity) to improve nasality prediction.

\end{abstract}

\vspace*{-5pt}
\section{Introduction}
\vspace*{-2.6pt}

Speech is produced by the coordinated movement of articulators such as tongue, velum, and lips that shape the acoustic signal produced by the larynx, forming alternations of vocal tract constriction (for consonants) and opening (for vowels) \cite{stevens2000acoustic}. These movement patterns can differ according to the language, dialect, abilities, and habits of the speaker, but the fact that the movements themselves overlap in time means that the evidence of their movement in the acoustic signal can be compressed, scattered across time, and sometimes obscured by co-occurring events. The result is that many linguistic phenomena that are hard to express in acoustic terms are more readily explained by differences in the timing and degree of vocal tract constriction \cite{CHO2009, Krivokapić_2014}. Systems that do speech inversion rely on ground truth articulatory variables; by using extracted acoustic features such as Mel Frequency Cepstral Coefficients (MFCCs), Mel-spectrograms, or the waveform itself as the input speech representation, the system can learn a mapping to the articulatory variables. However, none of the publicly available articulatory speech corpora have direct articulatory level data capturing the velar and glottal constrictions \cite{Westbury1994a, Tiede2017}. Therefore, most of the available SI systems (trained on these datasets) are limited to estimating the articulatory level information pertaining to lip and tongue constrictions \cite{illa18_interspeech, yashish_bigrnn, Shahrebabaki2020, siriwardena2022, udupa21_interspeech}.

Acoustic-to-articulatory speech inversion inspired by Articulatory Phonology \cite{Browman1992} maps the acoustic speech signal to the kinematic state of each constriction synergy (lips, tongue tip, tongue body, velum, and glottis) by its corresponding constriction degree and location coordinates, which are called vocal tract variables (TVs).  In this work, we extend a speech inversion system based on TVs to estimate the activity of the velar constriction by collecting a dataset that can be effectively used in training a speaker-independent SI system. We choose `Nasalance’ as the ground-truth to capture nasality for two reasons. First, it is a non-invasive measure and can be easily collected from a larger population, which will be beneficial in building a more generalizable, speaker-independent SI system. However, nasalance measures the ratio of acoustical energy between the nasal and oral tract. Accordingly, as a variable it is dependent on the amount of energy flowing through the glottis, and thus has only an indirect relationship with VP articulation \cite{Kochetov_2020, Rong2011}. Hence, the far reaching goal of the proposed SI system is not aimed at deriving aerodynamic relationships (such as nasalance) from the acoustic signal, but rather aimed at deriving VP articulatory movements. Our approach to achieve this goal was twofold. To investigate if nasalance is an accurate representation of velar constriction degree \cite{Browman1992}, we validated it with a more direct, invasive and accurate measure of VP activity called high-speed nasopharyngoscopy (HSN). To the best of our knowledge, this is the first time a SI system has been developed to estimate a proxy for a velar constriction degree TV, that will, in essence, capture the nasality in speech. 

The second reason for using nasalance derives from this susceptibility to glottal source effects. Learning a mapping from an acoustic representation that is rich with source level information (eg. Melspectrograms, auditory spectrograms) to nasalance, along with source features (eg. voicing and pitch) may positively influence the SI system performance for nasality prediction. To investigate such effects of using source features, Electroglottography (EGG) was synchronously collected to extract a voicing parameter, and aperiodicity, periodicity and pitch extracted from an aperiodicity, periodicity and pitch (APP) detector \cite{APPdetector} are also used as additional targets to further improve nasality prediction. 

The content in the following sections of the paper is organized as follows. In section \ref{sec:dataset}, we discuss the details of the dataset and explain the steps used to extract and validate the ground-truth nasalance parameter. In section \ref{sec:SI_system}, we highlight the details of the proposed SI system and the importance of using source features to estimate nasality. Finally in section \ref{sec:discussion}, we discuss the key conclusions drawn from the experiments and possible future directions.

\vspace*{-9pt}
\section{Dataset}
\label{sec:dataset}
\vspace*{-3pt}

This work is based on a subset of data from ongoing, collaborative data collection. The complete dataset, once collected, will be made public (subject to standard open source licensing agreements). One of the main goals of this dataset is to develop a speaker-independent speech inversion system to accurately estimate velar and glottal activity. The current dataset has been collected from 8 subjects (5 Female, 3 Male), and the demographic details of the speakers are listed in Table \ref{table:dataset}. 

\vspace*{-8pt}
\subsection{Ground-truth Nasalance Parameter}
\vspace*{-3pt}

\subsubsection{Background and Procedure for Data Collection}
\label{ssec:data_collection}
\vspace*{-5pt}

\vspace*{-5pt}
\begin{table}[th]
  \caption{Dataset Description. SW: South-west, C: Central, W: White, B: Black, H: Hispanic, NH : Non-Hispanic}
  \vspace*{-5pt}
  \centering
  \resizebox{\columnwidth}{!}{
  \begin{tabular}{c c c c c c}
    \toprule
    \textbf{Subject} &{\textbf{Gender}} &{\textbf{Language}} &{\textbf{HSN status}} &{\textbf{Age (years)}} &{\textbf{Ethnicity/Race}}\\
    \midrule
    1 &M & English(SWOhio) &HSN &28 &W, NH\\
    2 &F & English(STexas) &No HSN &24 &W, H\\
    3 &F & English(SWOhio) &No HSN &31 &W, NH\\
    4 &F & English(SWOhio) &No HSN &40 &W, NH\\
    5 &F & English(CKentucky) &No HSN &28 &B, NH\\
    6 &F & English(SWOhio) &HSN &34 &W, NH\\
    7 &M & English(SWOhio) &No HSN &23 &W, NH\\
    8 &M & English(SWOhio) &No HSN &35 &W, NH
    \\\bottomrule
    \vspace*{-5pt}
  \end{tabular}
  \label{table:dataset}}
  \vspace*{-8pt}
\end{table}

As noted above, nasalance is the relative proportion of nasal vs. oral acoustic output from two microphones (mic) mounted to the top and bottom of a separation plate located between the nose and upper lip to create an acoustic barrier. It is a simple, well-known, non-invasive and reliable technology for tracking VP constriction. We used a subset of speakers (subject 1 and subject 6) to synchronously collect a more direct but  invasive measure of VP constriction using high-speed nasopharyngoscopy (HSN). 

\begin{figure}[th]
    \centering
    \includegraphics[width=\linewidth, height=26mm]{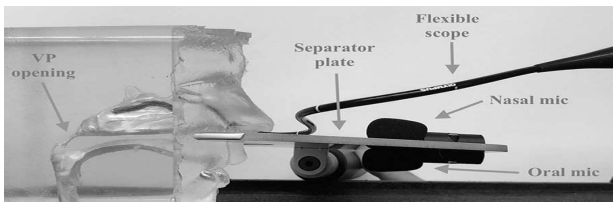}
    \caption{Illustration of the experimental setup. HSN measurements were taken by connecting a flexible scope to a high-speed video camera (not shown). The figure is taken from \cite{Oren_data_collection} in The Cleft Palate-Craniofacial.}
    \label{fig:data_collection_setup}
    \vspace*{-8pt}
\end{figure}

Figure \ref{fig:data_collection_setup} shows the setup used to collect the HSN and audio measurements to compute the nasalance parameter. For the HSN, a flexible scope (outer diameter: 2.2 or 3.6 mm) was connected to a video camera (MIRO 310; Vision Research, Inc., Wayne, New Jersey), and the  images were captured at a rate of 1000 frames/second using 304 $\times$ 256 pixel resolution. To collect the audio data, 2 microphones (1/4", Type 4958, Bruel and Kjær, Duluth, Georgia) were connected to the top and the bottom of the separation plate made of aluminum. Windscreens were used to cover the microphones to prevent interference from airflow directed toward the microphones. The separation plate was placed against the participant's upper lip to create an acoustic barrier between the oral and nasal audio recordings. The acoustic data from the microphones were captured at 51.2 kHz using a data acquisition system (NI 9234, National Instruments, Austin, Texas) and customized LabVIEW code that digitized and converted the data to a “.wav” audio file. The initiation of the audio recording and imaging data (from the HSV nasopharyngoscopy) was synchronized using an input/output module (NI 9402; National Instruments) \cite{Oren_data_collection}

Using this setup, approximately 10 minutes of speech material per subject was recorded.  This consisted of a mixture of short and long sentences and short paragraphs. For example, for nasality, the full set of prosodic nasal contrasts from Krakow et al. \cite{Krakow1988-dg} was included, including e.g. “hoe me” vs. “home E”, “seam ore” vs. “Seymour”. For voicing, sentences contrasting words such as “Dodd” vs. “Todd” in a carrier phase were included. Sentences illustrating consonant cluster articulatory patterns were drawn from Zsiga et al. \cite{ZSIGA1994, ZSIGA2004}. For cross-dataset comparison, we also included some sentences from speech materials used in the U.W. x-ray microbeam corpus \cite{Westbury1994a}.


\vspace*{-7pt}
\subsubsection{Nasalance Parameter}
\label{ssec:nasalane_parameter}
\vspace*{-3pt}

Oral and nasal mic signals collected from the nasometer set-up were used to compute the nasalance parameter. The baseline wander was first removed from the two signals using a high pass filter (cutoff around 0.1Hz). The Root Mean Square (RMS) signals were then computed for both oral and nasal signals separately. During the RMS signal generation, both the squared signals were smoothed out using a moving average filter with a window size of 1000 ($\sim$ 20 ms) samples. Then a nasalance parameter (Nasalance$_{raw}$) was computed using the equation \ref{eq_RMSE_raw} based on Bunton et al. \cite{Bunton2011-vw}. The Nasalance$_{raw}$ parameter was then downsampled to 100Hz and smoothed using a window of 10 samples (using Matlab function `Fastsmooth' by \cite{tom_2017}). The final nasalance parameter was then normalized to [-1,1] range to be used as the ground-truth for the speech inversion system.

\vspace*{-8pt}
\begin{equation}
  Nasalance_{raw} = \frac{RMS_{nasal}}{RMS_{nasal} + RMS_{oral}}
  \label{eq_RMSE_raw}
\end{equation}


\vspace*{-7pt}
\subsection{Validating Nasalance with HSN}
\label{ssec:correlation_analysis}
\vspace*{-3pt}

\begin{figure}[th]
    \centering
    \includegraphics[width=\linewidth, height=33mm]{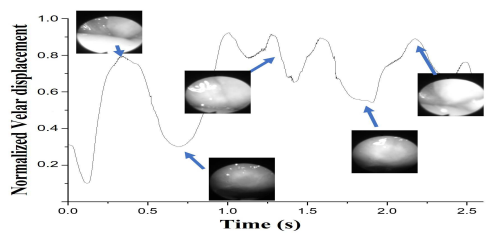}
    \caption{HSV intensity trace for a male native speaker of American English from Cincinnati, OH producing "It's a see more, Sid. It’s a seam ore, Sid”. Images of the VP port at key time points are indicated by arrows.}
    \vspace*{-4pt}
    \label{fig:HSV_trace}
\end{figure}

HSN was synchronously collected from subject 1 and subject 6 to assess the accuracy and agreement with the computed nasalance parameter. Here the temporal dynamics of the VP port is captured by summing the light intensity in the images (intensity of pixels) of the high-speed video (HSV) data. The resulting intensity trace has been shown to be an accurate measure for capturing the velum \cite{Oren_data_collection}. Figure \ref{fig:HSV_trace} shows a sample HSV intensity trace and the corresponding HSV images at different points in time. Here an open VP port would be overall characterized by darker regions that come from the cavity of the VP port. On the other hand, a closed VP port would be characterized with brighter regions because of the increased amount of light reflecting off the tissue. It should be noted that the HSN parameter shows a trough (i.e. lower values) for nasal sounds in speech.  This is in contrast to nasalance, which shows the opposite pattern of a peak.

\vspace*{-2pt}

The HSV data has a sampling rate of 1kHz and the nasalance parameter as discussed earlier is sampled at 100Hz. To match the number of samples to compute the cross correlations, the nasalance parameter is linearly interpolated to match with the HSV intensity trace. The Pearson correlation coefficients are then computed for each sample data from the subject. The average correlation coefficients across the samples for subject 1 and subject 6 are -0.6081(\emph{p$<$0.001}) and -0.5136(\emph{p$<$0.001}) respectively. These statistically significant negative correlations give an important validation on the accuracy of the computed nasalance parameter with respect to HSN.

\begin{table*}
  \caption{PPMC scores (mean and .std across 8 trials) for the SI systems trained with and without source features as additional targets to estimate nasalance.}
  \vspace*{-5pt}
  \centering
  \resizebox{\textwidth}{!}{
  \begin{tabular}{c c c c c c c}
    \toprule
    \textbf{} &{\textbf{Nasalance}} &{\textbf{Voicing}} &{\textbf{Perio.}} &{\textbf{Aperio.}} &{\textbf{Pitch}} &{\textbf{Average}}\\
    \midrule
    SI-SF &\textbf{0.7341(0.02)} &0.80541(0.01) &0.9008(0.03) &0.8257(0.02)&0.7995(0.03) &0.8131(0.03)\\
    SI-noSF &0.6967(0.02) &- &- &- &- &-
    \\\bottomrule
  \end{tabular}
  \label{table:ppmc_scores}
  \vspace*{-5pt}
  }
\end{table*}

\vspace*{-7pt}
\subsection{Patterns of timing for Nasality}
\label{ssec:timing_analysis}
\vspace*{-4pt}

\begin{figure}[th]
    \centering
    \includegraphics[width=\linewidth, height=85mm]{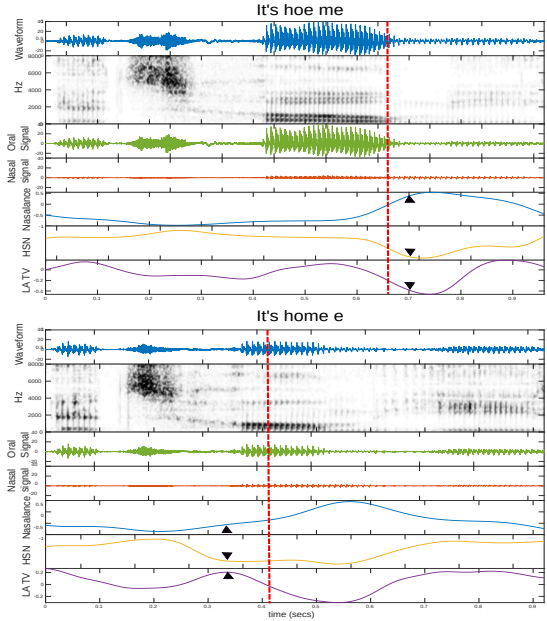}
    \vspace*{-14pt}
    \caption{The vertical red dash lines in the top and bottom panels mark the onset of bilabial contact for the /m/. The black triangles mark velum lowering offset and the coordinated event in the lower lip (lip raising onset or offset)}
    \label{fig:timing_plot}
    \vspace*{-5pt}
\end{figure}
\vspace*{-5pt}

A number of studies have shown that American English shows different patterns of velum raising and lowering (i.e. VP port constriction) according to syllabic organization \cite{KRAKOW1999}. As shown in Krakow et al. \cite{KRAKOW1999} an example of this pattern for “home E” vs “hoe me” is that the velum moves earlier and the VP port stays open longer when the /m/ is in the rime (home) than when the /m/ is in the onset
of the following word (me). The lip-velum coordination during the syllable-initial and -final nasal was also observed in Krakow et al. \cite{KRAKOW1999}, where it has been noted that there is close temporal proximity between the end of velum lowering and the beginning of lip raising for the syllable-initial and a large offset between the end of velum lowering and the end of lip raising for syllable-final.  

To see if the nasalance parameter will also showcase such patterns (word-initial vs word-final /m/) with respect to the HSN and lip movement, the words `hoe me' and `home e' were analyzed. Figure \ref{fig:timing_plot} shows the data for `It's hoe me' and `It's home e' collected from subject 1 in the dataset. To analyze the lip movement pattern, the lip aperture tract variable (LA TV) was extracted from the articulatory speech inversion system in \cite{siriwardena2022}. Both the HSN and nasalance patterns shown in Figure \ref{fig:timing_plot} replicate the timing patterns described in Krakow et al.\cite{KRAKOW1999} with respect to the LA TV. Data from a larger group of subjects is needed to further verify the pattern.


\vspace*{-7pt}
\subsection{Voicing parameter: EGG envelope}
\vspace*{-3pt}

Electroglottography (EGG) is a well-established technology for tracking vocal fold oscillation, using the degree of electrical conductance across the glottal gap between electrodes placed on the two parallel outer sides of the throat. In this study, EGG data was also collected (from all the subjects) synchronously with the other HSN and audio measurements in section \ref{ssec:data_collection}. 

The EGG signal is sampled at 51.2 KHz, and to compute a parameter which can capture the voicing activity of speech, the envelope of the EGG signal was extracted. As with the nasalance parameter, we first high pass filtered the signal to remove the baseline wander. Then the magnitude of the Hilbert transform \cite{FELDMAN2001642} was computed as the envelope of the EGG signal. The envelope was downsampled to 100 Hz and smoothed and normalized the same way to the nasalance parameter to generate the final voicing parameter. 

\vspace*{-7pt}
\section{Speech Inversion System}
\label{sec:SI_system}
\vspace*{-2pt}

\subsection{Input Audio Representation}
\vspace*{-3pt}

The audio recorded by the oral and nasal mic signals were mixed together to create a combined audio signal. The combined signal was then downsampled to 16kHz and segmented to 2 second long segments. The shorter, remaining segments were zero padded at the end. The segmentation was done mainly to increase the number of audio samples to train the DNN based SI system and to have input acoustic representations of fixed dimensionality to the input layer of the DNN model.

We used auditory spectrograms (Audspec) \cite{Wang_auditory_spec} as the input speech representation for the SI system. The auditory spectrograms have a logarithmic frequency scale and provide a unified multi-resolution representation of the spectral and temporal features likely critical in the perception of sound \cite{Wang_auditory_spec}.

\vspace*{-7pt}
\subsection{Model Architecture and Training}

\vspace*{-3pt}
\subsubsection{Model Architecture}
\vspace*{-3pt}

\begin{figure}[th]
    \centering
    \includegraphics[width=\linewidth, height = 33mm]{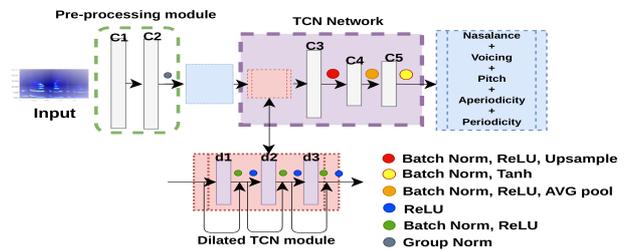}
    \caption{Model architecture (SI system). Here C1-C5 represent 1D-CNN layers and d1-d3 represent 1D dilated CNN layers}
    \vspace*{-5pt}
    \label{fig:model_archi}
\end{figure}

We developed a Temporal Convolution Network (TCN) based SI system inspired by the work in \cite{siriwardena2022}. The model was optimized using the Mean Squared Error (MSE) loss computed between the predicted parameters and the ground truth. The SI system was implemented in PyTorch with 1-D convolutional (CNN) layers. Figure \ref{fig:model_archi} shows the proposed model architecture with its sub-modules used for pre-processing and dilated TCN. The pre-processing module contains two 1-D CNN layers with 1$\times$1 kernels (C1, and C2), which have 128 filters each. The d1, d2 and d3 dilated CNN layers have a kernel size of 3 with 1,4 and 16 dilation rates respectively. Upsampling (window size 4) was done after C4 layer and average pooling (window size 5) was done after C5 layer along with BatchNorm layers after every CNN layer in the TCN network. The upsampling and average pooling operations take care of matching the time dimension of the input spectrograms to the target time dimension of TVs.

\vspace*{-8pt}
\subsubsection{Model Training}
\vspace*{-3pt}

All the model parameters were randomly initialized with a seed (=7) for reproducibility. Table \ref{table:hyperparmeters} lists the hyper-parameters and the corresponding values considered to fine-tune the model. A grid search was performed when fine-tuning the hyper-parameters and the best parameters were chosen based on the validation loss. All the models were implemented with PyTorch machine learning framework and trained with NVIDIA TITAN X GPUs. The best performing model has around ~1 million trainable parameters, takes around 8 minutes ($\pm2$) to converge, and can be found in a Github repository\footnote[1]{https://github.com/Yashish92/TCN-SI-tool-Nasality} 

\begin{table}
  \caption{Hyperparameter Tuning for the TCN model}
  \vspace*{-5pt}
  \centering
  \resizebox{\columnwidth}{!}{
  \begin{tabular}{l c c}
    \toprule
    {\textbf{Parameter}} &{\textbf{Possible Values}} &{\textbf{Chosen Values}}\\
    \midrule
    Learning Rate &[1e-4, 3e-4, 1e-3, 1e-2] &1e-3\\
    Batch size &[16,32,64,128] &64\\
    Optimizer &ADAM, RMSprop, SGD  &ADAM\\
    Rate scheduler & ExponentialLR, PolynomialLR &ExponentialLR
    \\\bottomrule
  \end{tabular}
  \label{table:hyperparmeters}}
  \vspace*{-12pt}
\end{table}

The dataset was divided into training, validation and testing splits, so that the training set has utterances from 6 speakers (4 females, 2 males). The validation and testing splits have data from 2 speakers (1 male, 1 female) with 1/2 of the data from each speaker in the validation split and the other half in the test split. None of the data from the speakers in the validation and test splits were included in the training split and hence all the models are trained in a `speaker-independent' fashion. The splits also ensured that around 70\% of the total number of utterances were present in training (~1 hour of speech), and all the allocations were done in a completely random manner. 




\vspace*{-7pt}
\subsection{Results of Speaker-independent Speech Inversion}
\vspace*{-3pt}

Two speech inversion systems were trained to estimate the nasalance parameter from the input auditory spectrograms. Pearson Product Moment Correlation (PPMC) score is used as the metric to evaluate the predictions by the SI systems. Table \ref{table:ppmc_scores} shows the PPMC scores for correlations between the estimated and ground-truth nasalance parameter for the systems trained with additional source features as targets (SI-SF) and the one with nasalance parameter as the only target (SI-noSF). 

\begin{figure}[th]
    \centering
    \includegraphics[width=\linewidth, height=56mm]{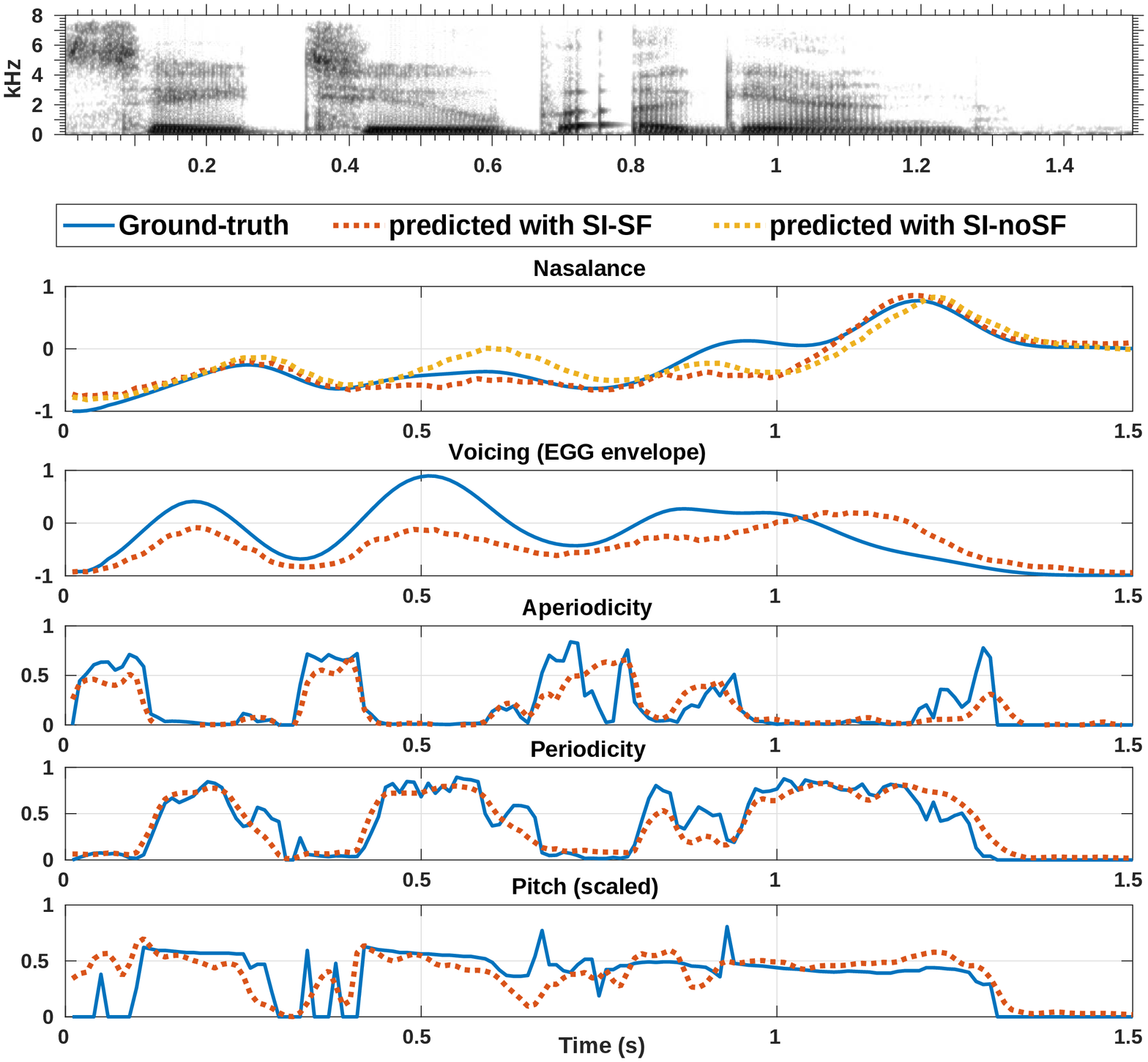}
    \caption{Nasalance and source features for the utterance ‘Say tube again’ estimated by the SI-SF model and nasalance estimated by the SI-noSF model with respect to the ground-truth . Solid blue Line - ground truth, red dotted line - predictions by the SI-SF, yellow dotted Line - predictions by SI-noSF.}
    \vspace*{-18pt}
    \label{fig:si_estimations}
\end{figure}

Figure \ref{fig:si_estimations} shows sample nasalance estimation by the SI-SF and SI-noSF models for an utterance in the test set. The utterance, `Say tube again' contains a nasal consonant \textipa{[n]} around 1.15-1.25 seconds which is captured by both the SI systems. However, it is important to note that the nasalance parameter estimated by the SI-SF model has better agreement with the ground-truth compared to the SI-noSF model. 

\vspace*{-8pt}
\section{Discussion and Conclusion}
\label{sec:discussion}
\vspace*{-2pt}

The results of correlation analysis in the section \ref{ssec:correlation_analysis}
gives a general, but an important validation for the nasalance parameter with respect to the more direct HSV intensity trace. The fact that we found known patterns of timing for nasality (discussed in section \ref{ssec:timing_analysis}) further supports the validity of using nasalance as a proxy variable for velopharyngeal constriction. This work highlights the performance of our SI system in estimating velopharyngeal movement dynamics for unseen speaker data. It also shows that incorporating source features as additional targets improves the estimation accuracy of the velopharyngeal movement parameter. This is consistent with the observations made in \cite{siriwardena2022} with conventional acoustic-to-articulatory speech inversion, and could also suggest that the TCN model is particularly sensitive to source/VP interactions.


In future work, the authors plan to improve the performance and generalizability of the current SI system by training on data from a larger group of subjects (from the ongoing data collection). More emphasis will also be made on validating and fine tuning the nasalance parameter as a proxy to the velar TV. Further experiments will also be done to understand what the DNN models are actually picking as source-filter interactions that are ultimately helping the overall SI task. 

To summarize, in this work we present the details on a dataset collected to estimate the velar and glottal activity in speech. We particularly looked into estimating a validated nasalance parameter (as a proxy to a velar TV) using a speaker-independent SI system. It should be noted, that having a SI system to estimate parameters directly related to the velar (and glottal) constrictions can be hugely beneficial, since it gives an almost complete articulatory level representation of speech which can be useful in diverse speech applications (eg. articulatory speech synthesis \cite{wu22i_interspeech, siriwardena_mirrorNet}). An accurate, validated speech inversion system would also be a significant breakthrough for researchers with little or no ability to collect articulatory data directly, e.g. scholars without well-equipped phonetics laboratories, scholars doing field studies in dispersed communities. While speech inversion data is not equivalent to direct observation, it may enable hypothesis formation and testing that will motivate more targeted studies. 

\vspace{-6pt}
\section{Acknowledgement}
\vspace{-2pt}

This work was supported by the NSF grant BCS2141413

\vspace{-6pt}
\bibliographystyle{IEEEtran}
\bibliography{mybib}

\end{document}